\documentclass[pra,twocolumn,showpacs,amsmath,amssymb]{revtex4}
\usepackage{graphicx}
\usepackage{bm}
\usepackage{amsmath,amssymb}

\begin{document}

\title{Quantum computational tensor network on 
string-net condensate}  

\author{Tomoyuki Morimae}
\affiliation{
Universit\'e Paris-Est Marne-la-Vall\'ee,
77454 Marne-la-Vall\'ee Cedex 2, France
}
\date{\today}
            
\begin{abstract}
String-net condensate is a new class of materials which exhibits
quantum topological order.
In order to answer the important question, 
``how useful is string-net condensate in quantum information
processing?",
we consider the simplest example of string-net condensate,
namely
the $Z_2$ gauge string-net condensate on the 
two-dimensional hexagonal 
lattice, 
and investigate possibilities of
universal measurement-based quantum computation
on it by using the framework of quantum computational
tensor network.
We show that universal measurement-based quantum computation is possible
by coupling two correlation space wires
with a physical two-body interaction. 
We also show that
universal measurement-based quantum computation is possible solely
with single-qubit measurements if we consider
a slightly modified version of the $Z_2$ gauge string-net condensate
on the two-dimensional hexagonal lattice.
These results suggest that 
even the
simplest example of string-net condensate 
is equipped with
the correlation space
that has the capacity for universal 
quantum computation.
\end{abstract}
\maketitle  
\section{Introduction}
Exploiting the full 
potential of quantum many-body states
for quantum information processing
is one of the most central research subjects in 
today's quantum information
science and condensed matter physics.
Plenty of theoretical and experimental
researches over the last few years have contributed to
the important cross-fertilization between those two fields.
In particular, the seminal work by Raussendorf and Briegel
about the one-way
quantum computation~\cite{Raussendorf} has offered 
the clear view to the role of quantum many-body states played in  
quantum computation.
Once the special many-body state, which is called the cluster state,
is prepared, universal quantum computation can be performed
by adaptively measuring each qubit.
Recently, the one-way quantum computation was generalized
to more abstract and beautiful framework,
so called quantum computational 
tensor network~\cite{Gross1,Gross2,Gross3}.
This framework, which cleverly uses the  
matrix product representation~\cite{Fannes,Garcia}
(or, more generally, the tensor-network representation~\cite{MPS_review1,MPS_review2}) of 
quantum many-body states,
has established the novel concept of quantum computation,
namely {\it virtual quantum computation in the correlation 
space}.
The exploration of resource many-body states has been thus replaced
with the exploration of correlation spaces which have full capacity for universal
quantum computation~\cite{Cai_magnet,Miyake_AKLT,Chen_tricluster,Cai_CS,Morimae_upload,
Morimae_blind,MF,FM,Miyake_2dAKLT,reduction,Darmawan,experiment1,experiment2}.



On the other hand, in quantum many-body physics, such as condensed matter
physics, statistical physics, and nuclear physics,
orders and phase transitions lie at the heart of main research topics.
Traditionally, phase transitions have been studied in the framework of 
Landau's symmetry-breaking theory~\cite{Landau}.
However, a new type of order, so called quantum topological order, 
has been discovered in several many-body systems,
such as 
the fractional quantum Hall system~\cite{Hall},
and it has been pointed out that
the quantum topological order slips through
the framework of Landau's symmetry-breaking theory.
String-net condensate is 
a new class of materials introduced in Ref.~\cite{Wen} to describe
such a quantum topological order.
As is pointed out in Ref.~\cite{WenPTP}, 
it is of great worth to explore applications of such a new
type of condensate to technologies, since
there have been many successful examples in the applications of
Landau's symmetry-breaking states 
to technologies, such as
a hard disk drive, a liquid crystal display, and
semiconducting devices.
In particular, it is very important to answer the question,
``how useful is string-net condensate in quantum information processing?".
Indeed, the famous scheme of the fault-tolerant quantum memory with
the Kitaev's toric code state~\cite{toric1,toric2},
which is the $Z_2$ gauge string-net condensate on the two-dimensional square
lattice,
would be the first great example of an application of string-net condensate
to quantum information technology.

In this paper, by using the framework of 
quantum computational tensor network~\cite{Gross1,Gross2,Gross3},
we study how useful is string-net condensate in 
quantum computation.
For this purpose,
we consider the simplest example of string-net condensate,
namely 
the $Z_2$ gauge string-net condensate
on the two-dimensional hexagonal lattice.
As is shown in Ref.~\cite{BR}, Kitaev's toric code state
on the
two-dimensional square lattice is not a universal resource state in the sense
that the measurement-based quantum computation solely with
single-qubit measurements on that state can be classically
simulated.
Their result can be generalized to the two-dimensional hexagonal 
lattice~\cite{generalize}.
Therefore, the $Z_2$ gauge
string-net condensate on the two-dimensional 
hexagonal
lattice is neither a universal resource state in the strict sense 
(here, the strict definition of the universality is that universal 
measurement-based quantum computation can be performed solely with
single-qudit measurements.)
However, we show that universal measurement-based quantum computation
is possible on the $Z_2$ gauge string-net condensate
on the two-dimensional hexagonal lattice,
by coupling two correlation space wires with 
a physical two-body interaction.
Considering the fact that string-net condensate is
central in
condensed-matter physics and the fact that many examples
of resource states which require
physical two-body interactions
have greatly contributed to
the recent development of measurement-based quantum 
computation~\cite{Gross3,Miyake_AKLT,Darmawan,Miyakeholographic,Morimae_blind,Bartlett,experiment1,experiment2}, it is still very important to investigate
possibilities of measurement-based quantum computation with
the help of physical two-body interactions.
We also show that a 
slightly modified version of the $Z_2$ gauge string-net condensate
on the two-dimensional hexagonal lattice 
is a universal resource state in the strict sense.
Our results imply that even the simplest example of 
string-net condensate 
is equipped with the correlation space that has
the capacity for universal
quantum computation.

Surprisingly, the possibility of universal measurement-based
quantum computation on string-net condensate was an open problem
even if a physical two-body interaction is allowed.
The cluster state~\cite{Raussendorf} is 
not topologically ordered, since some quasi-local unitary operations
can change the cluster state 
into a product state~\cite{LU,Yoshida,Hastings}.
In Refs.~\cite{Gross1,Gross2},
the measurement-based quantum computation on the slightly
modified version of the Kitaev's toric code state
was studied.
However, it is not clear whether that resource state
is string-net condensate or not.
The two-dimensional AKLT state~\cite{Miyake_2dAKLT}
and the resource state proposed in Ref.~\cite{Cai_magnet}
do not seem to be string-net condensate at least in the $Z$ basis
from their tensor-network structures~\cite{Miyake_2dAKLT,Cai_magnet}.
Furthermore, since these resource states are not qubit states but qudit states,
string-net condensate should be more complicated than $Z_2$ gauge type
if any.
In Ref.~\cite{BR},
a string-net condensate on a highly non-local graph 
(Fig.4 (b) of Ref.~\cite{BR}),
which can be converted into a cluster
state by single-qubit measurements and hence universal, was proposed.
However, it was an open problem whether such a strong assumption
(i.e., the highly non-local graph) can be weakened or not. 
Our results answer the question:
we can perform universal measurement-based quantum
computation on a string-net condensate
without considering highly non-local graph structure,
if two correlation space wires are coupled with
a physical two-body interaction,
or if the coefficients of the linear-combination
of closed-loop configurations are slightly modified.


\section{
$Z_2$ gauge string-net condensate}
\begin{figure}[htbp]
\begin{center}
\includegraphics[width=0.35\textwidth]{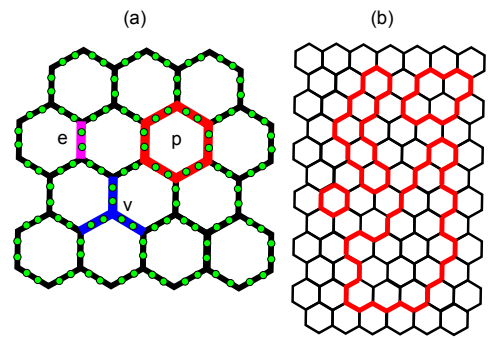}
\end{center}
\caption{(Color online.) 
(a): The two-dimensional hexagonal lattice. Green circles represent qubits.
Two qubits are placed on each edge.
The red hexagon indicates the set $S_p$ of 12 qubits in the
plaquette $p$. The blue tree
indicates the set $S_v$ of six qubits in the three edges
associated with the vertex $v$.
The purple bond indicates the set $S_e$ of two qubits in the edge $e$.
(b): An example of the string-net configuration where only closed loops exist.
Red bonds represent strings. 
} 
\label{hexfig1}
\end{figure}

\begin{figure}[htbp]
\begin{center}
\includegraphics[width=0.35\textwidth]{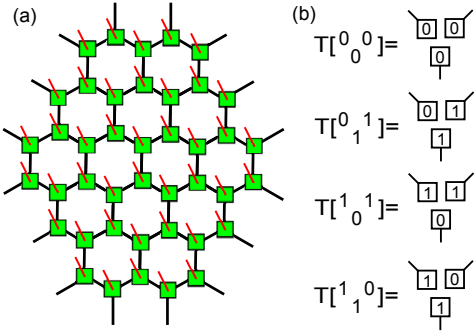}
\end{center}
\caption{(Color online.) 
(a): The tensor-network representation of the ground state $|G\rangle$. Green
boxes represent tensors $T$. Red lines represent physical legs.
(For simplicity, a single red line represents three physical legs).
Black lines are virtual legs.
(b): The tensor $T$. 
It has three physical legs, each of
which corresponds to a single qubit,
and three virtual legs, each of which corresponds to
the two-dimensional Hilbert space.
} 
\label{hexfig2}
\end{figure}

Let us consider the two-dimensional hexagonal lattice where two qubits
are placed on each edge (Fig.~\ref{hexfig1} (a)).
We define the 
Hamiltonian~\cite{tensor_representation1,tensor_representation2} $H$ on this lattice by
$
H\equiv-\sum_p\bigotimes_{i\in S_p}X_i
-\sum_v\bigotimes_{i\in S_v}Z_i
-\sum_e\bigotimes_{i\in S_e} Z_i,
$
where $X$ and $Z$ are Pauli operators.
$p$ indicates a hexagonal plaquette, 
$v$ indicates a vertex,
and
$e$ indicates an edge (see Fig.~\ref{hexfig1} (a)).
$S_p$ is the set of 12 qubits in the plaquette $p$,
$S_v$ is the set of six qubits in the three edges associated with the vertex $v$,
and $S_e$ is the set of two qubits in the edge $e$
(see Fig.~\ref{hexfig1} (a)).
The two-body interaction $-\bigotimes_{i\in S_e}Z_i$
forces two qubits on the edge $e$ to be
the up-up state $|0\rangle\otimes|0\rangle$ 
or the down-down state $|1\rangle\otimes|1\rangle$.
Here, $Z|s\rangle=(-1)^s|s\rangle$ $(s\in \{0,1\})$.
If two qubits on an edge are $|1\rangle\otimes|1\rangle$, we consider that the
edge is occupied by a string.
On the other hand, if the two spins are $|0\rangle\otimes|0\rangle$,
the edge is considered to be vacant (see Fig.~\ref{hexfig1} (b)).
The six-body interaction $-\bigotimes_{i\in S_v}Z_i$ forces
strings to form closed loops (see Fig.~\ref{hexfig1} (b)).
The twelve-body interaction 
$-\bigotimes_{i\in S_p}X_i$ works as the kinetic term for
such closed loops.
Therefore, the ground state $|G\rangle$ of the Hamiltonian $H$ is 
the equal weight superposition of
all closed loop configurations. 
Such a state is called 
the string-net condensate~\cite{Wen}.
It is known~\cite{tensor_representation1,tensor_representation2} that the ground state $|G\rangle$ has a 
simple tensor-network representation.
As is shown in Fig.~\ref{hexfig2} (a),
the tensor $T$ defined in Fig.~\ref{hexfig2} (b)
is placed on each vertex of the hexagonal lattice,
and its virtual legs are connected with three nearest-neighbor tensors.
As is shown in
Fig.~\ref{hexfig2} (b),
the tensor $T$ has three
physical legs, each of which corresponds to a single qubit,
and three virtual legs, each of which corresponds to the
two-dimensional Hilbert space.

\begin{figure}[htbp]
\begin{center}
\includegraphics[width=0.25\textwidth]{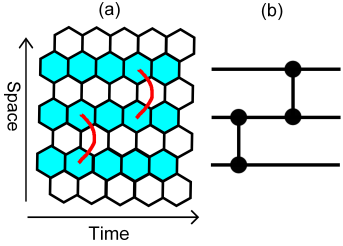}
\end{center}
\caption{(Color online.) 
(a): A schematic illustration of our measurement-based quantum
computation. 
(b): The corresponding quantum circuit.
The horizontal lines are single-qubit wires, whereas the vertical
lines are CZ gates.
} 
\label{hexfig3}
\end{figure}

In Fig.~\ref{hexfig3}, 
we illustrate how our measurement-based quantum computation runs on
the ground state $|G\rangle$. 
As is shown in Fig.~\ref{hexfig3} (a),
each horizontal line of plaquettes encodes a 
logical single-qubit wire, 
and
the Controlled-Z (CZ) gate is implemented by interacting two
nearest-neighbour logical single-qubit wires.
Figure~\ref{hexfig3} (b) shows the quantum circuit
which corresponds to Fig.~\ref{hexfig3} (a).
Throughout this paper, we will adopt
the framework of quantum 
computational tensor network~\cite{Gross1,Gross2,Gross3}. Readers who are
not familiar with that framework can refer to 
Refs.~\cite{Gross1,Gross2,Gross3,Cai_magnet,
Cai_CS,Morimae_upload,Morimae_blind,
MF,FM,Miyake_AKLT,Miyake_2dAKLT,reduction}.

\section{Implementation of gates}

\begin{figure}[htbp]
\begin{center}
\includegraphics[width=0.45\textwidth]{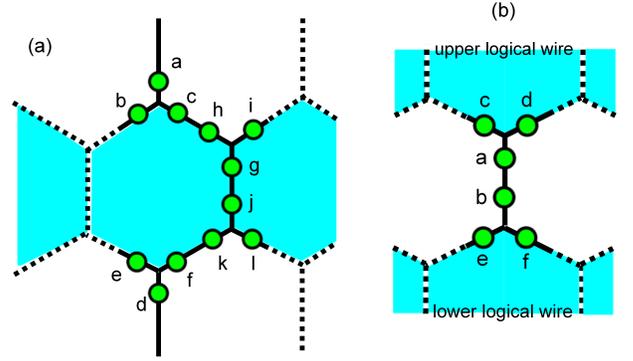}
\end{center}
\caption{(Color online.) 
(a) A single-qubit rotation.
(b) The CZ gate.
} 
\label{gate}
\end{figure}

Let us see Fig.~\ref{gate} (a).
If we perform the $Z$ measurement on site $a$ and
the $X$ measurement on sites $b$ and $c$, 
the operation $X^{\mu_a}Z^{\mu_b\oplus \mu_c}$ is implemented in the path 
$b\to c$.
Here, $\mu_s\in\{0,1\}$ ($s=a,b,c$) is the measurement result
($\mu_s=0$ corresponds to the result of the
eigenvalue +1 and $\mu_s=1$ corresponds to the result of the eigenvalue
-1, respectively).
In the same way, if we perform the $Z$ measurement on site $d$ and
the $X$ measurement on sites $e$ and $f$, 
the operation $X^{\mu_d}Z^{\mu_e\oplus \mu_f}$ is implemented in the path $e\to f$.
Let $\{|0\rangle_u,|1\rangle_u\}$ 
($\{|0\rangle_l,|1\rangle_l\}$) 
be the basis of the qubit that flows the upper (lower) path,
$...\to b\to c\to h\to i\to...$  
($...\to e\to f\to k\to l\to...$).
We encode a single logical qubit $\{|0_L\rangle,|1_L\rangle\}$ 
by using these two qubits~\cite{single_can_be_done}.
Let us define two types of the encoding~\cite{two_encoding}:
the type-I encoding is defined by
$|0_L^{I}\rangle=|0\rangle_u\otimes|0\rangle_l$
and
$|1_L^{I}\rangle=|1\rangle_u\otimes|1\rangle_l$,
whereas
the type-II encoding is defined by
$|0_L^{II}\rangle=|0\rangle_u\otimes|1\rangle_l$
and
$|1_L^{II}\rangle=|1\rangle_u\otimes|0\rangle_l$.
Then, 
the effect of the operation
$(X^{\mu_a}Z^{\mu_b\oplus\mu_c})\otimes
(X^{\mu_d}Z^{\mu_e\oplus\mu_f})$
is just the change of the encoding type,
the logical $X$ operation, or the logical $Z$ operation.
This is understood as follows:
Let 
$E\equiv(X^{\mu_a}Z^{\mu_b\oplus\mu_c})\otimes(X^{\mu_d}Z^{\mu_e\oplus\mu_f})$.
Then,
by using the facts that $(X\otimes X)|0_L^I\rangle=|1_L^I\rangle$
and that $XZ=-ZX$,
we obtain
$E(\alpha|0^I_L\rangle+\beta|1^I_L\rangle)
=
\alpha E|0^I_L\rangle+(-1)^r\beta(X\otimes X)E|0^I_L\rangle$,
where $r=\mu_b\oplus\mu_c\oplus\mu_e\oplus\mu_f$.
Note that $E|0^I_L\rangle=(X\otimes X)^v|0^{I/II}_L\rangle$
up to a phase factor
with some $v\in\{0,1\}$, where ``$I/II$" stands for ``type-I encoding
or type-II encoding".
Therefore,
\begin{eqnarray*}
E\Big(\alpha|0^I_L\rangle+\beta|1^I_L\rangle\Big)
&=&X^v_L\Big(\alpha|0^{I/II}_L\rangle+(-1)^r\beta|1^{I/II}_L\rangle\Big)\\
&=&X^v_LZ^r_L
\Big(\alpha|0^{I/II}_L\rangle+\beta|1^{I/II}_L\rangle\Big),
\end{eqnarray*}
where $X_L=X\otimes X$ and $Z_L=Z\otimes I$.
Similar result is obtained for $\alpha|0^{II}_L\rangle+\beta|1^{II}_L\rangle$.

In order to perform the logical single-qubit $z$-rotation
$e^{-iZ_L\theta/2}$,
we do the $X$ measurement
on site $i$, $k$, and $l$,
the $Z$ measurement on site $g$, 
and the measurement in the basis
$\{\frac{1}{\sqrt{2}}(|0\rangle\pm e^{-i\theta}|1\rangle) 
\}$ 
on site $h$.  
Then, the operation
$
[
(Z^{\mu_h\oplus \mu_i}X^{\mu_g})\otimes
(Z^{\mu_k\oplus \mu_l}X^{\mu_g})
]
(e^{-iZ\theta/2}\otimes I)
$
is implemented up to a phase factor.
This is logical single-qubit $z$-rotation up to some logical Pauli byproducts.
(If there is the logical $X$ byproduct $X_L=X\otimes X$ before
this $z$-rotation, $\theta$ should be replaced with $-\theta$.)

In order to perform
the logical single-qubit $x$-rotation $e^{-iX_L\theta/2}$,
we do
the $X$ measurement on site $h$, $i$, $k$, and $l$,
and the measurement in the basis
$\{\cos\frac{\theta}{2}|0\rangle+i\sin\frac{\theta}{2}|1\rangle,
\sin\frac{\theta}{2}|0\rangle-i\cos\frac{\theta}{2}|1\rangle\}$ on
site $g$. 
Then, the operation
$
[
(Z^{\mu_h\oplus \mu_i}X^{\mu_g})
\otimes (Z^{\mu_k\oplus \mu_l}X^{\mu_g})
]
(\cos\frac{\theta}{2}I\otimes I-i\sin\frac{\theta}{2}X\otimes X)
$
is implemented up to the phase factor.
This is logical single-qubit $x$-rotation up to some logical Pauli
byproducts.
(If there is the logical $Z$ byproduct $Z_L=Z\otimes I$ before
this $x$-rotation, $\theta$ should be replaced with $-\theta$.)

Let us see Fig.~\ref{gate} (b).
In order to perform the logical CZ gate between the upper logical wire
and the lower logical wire,
we use the coupling technique~\cite{Gross1,Gross2,Gross3,Miyake_AKLT}.
We first apply the physical CZ interaction
between sites 
$c$ and $e$.
We next do the $Z$ measurement on site $a$,
and the $X$ measurement on sites $c$, $d$, $e$,
and $f$.
Then, the operation 
$(Z^{\mu_c\oplus\mu_d}X^{\mu_a})
\otimes 
(Z^{\mu_e\oplus\mu_f}X^{\mu_a})
(|00\rangle\langle00|+|01\rangle\langle01|+|10\rangle\langle10|
-|11\rangle\langle11|)$
up to a phase factor
is implemented.
If the upper logical wire is in the type-I encoding, this is
the logical CZ gate up to some logical Pauli byproducts.
If the upper logical wire is in the type-II encoding, this is
the logical CZ gate plus $I_L\otimes Z_L$ 
up to some logical Pauli byproducts.

Note that these physical CZ interactions can be also done in advance before
starting the measurement-based quantum computation.
In other words, we can start with the resource state
$|G'\rangle\equiv\bigotimes_{i,j} CZ_{i,j}|G\rangle$,
where physical $CZ$ interactions are periodically
applied on appropriate places.
In this case, the entire computation can be done solely with only
single-qubit measurements,
since an unwanted CZ gate can be canceled by 
doing the identity operation (plus some Pauli byproducts)
until we arrive at the next CZ gate which cancels the previous 
one~\cite{Morimae_blind}.
Furthermore, $|G'\rangle$ is still string-net condensate,
although the coefficient of the linear combination
of closed-loop configurations are more complicated:
$
|G'\rangle=\sum_{\xi}(-1)^{f(\xi)}|\xi\rangle,
$
where $\xi$ is a closed-loop configuration
and $f(\xi)$ is a certain function of $\xi$.

Finally, let us consider the initialization and the final readout
of logical qubits.
A virtual leg of the tensor $T$ is initialized 
in the computational basis if we do the physical $Z$ measurement 
on two physical legs that does not correspond to
the virtual leg that we want to initialize. 
For example, in Fig.~\ref{gate} (a),
if we do the $Z$ measurement on
sites $h$ and $g$,
the site $i$ is initialized as $|\mu_h\oplus\mu_g\rangle$.
Thus the initialization is possible.
The final readout in the computational basis is also possible in the
similar way~\cite{finite_size_effect}.


\section{Conclusion}
In this paper, we have shown that the correlation space of the $Z_2$ gauge
string-net condensate
on the two-dimensional hexagonal lattice has sufficient capacity for universal measurement-based quantum
computation. 

Since the Hamiltonian considered in this paper is gapped and 
frustration-free, our quantum computation 
can enjoy the energy-gap protection by adiabatically turning off appropriate
interactions as in Refs.~\cite{Miyake_AKLT,Miyake_2dAKLT}.
It would be interesting to investigate whether
our model  
has 
the robustness of the computational
capacity against slight changes of Hamiltonian parameters,
the edge-state picture, and
the symmetrically protected order~\cite{Miyakeholographic}.

\acknowledgements
The author is supported by ANR, Stat Quant (JC07 07205763).



\begin{thebibliography}{00}
\bibitem{Raussendorf}
R. Raussendorf and H. J. Briegel, Phys. Rev. Lett. {\bf86}, 5188 (2001).
\bibitem{Gross1}
D. Gross and J. Eisert, Phys. Rev. Lett. {\bf98}, 220503 (2007).
\bibitem{Gross2}
D. Gross, J. Eisert, N. Schuch, and D. Perez-Garcia,
Phys. Rev. A {\bf76}, 052315 (2007).
\bibitem{Gross3}
D. Gross and J. Eisert, Phys. Rev. A {\bf82}, 040303(R) (2010).
\bibitem{Fannes}
M. Fannes, B. Nachtergaele, and R. F. Werner, J. Phys. A {\bf24}, L185 (1991).
\bibitem{Garcia}
D. Perez-Garcia, F. Verstraete, M. M. Wolf, and J. I. Cirac,
Quant. Inf. Comput. {\bf7}, 401 (2007).
\bibitem{MPS_review1}
F. Verstraete, J. I. Cirac, and V. Murg,
Adv. Phys. {\bf57}, 143 (2008).
\bibitem{MPS_review2}
J. I. Cirac and F. Verstraete,
J. Phys. A: Math. Theor. {\bf42}, 504004 (2009).

\bibitem{Miyake_AKLT}
G. K. Brennen and A. Miyake, Phys. Rev. Lett. {\bf101}, 010502 (2008).

\bibitem{Cai_magnet}
J. Cai, A. Miyake, W. D\"ur, and H. J. Briegel, 
Phys. Rev. A {\bf82}, 052309 (2010).


\bibitem{Cai_CS}
J. Cai, W. D\"ur, M. Van den Nest, A. Miyake, and H. J. Briegel,
Phys. Rev. Lett. {\bf103}, 050503 (2009).

\bibitem{Miyake_2dAKLT}
A. Miyake, Ann. Phys. {\bf326}, 1656 (2011).

\bibitem{reduction}
X. Chen, R. Duan, Z. Ji, and B. Zeng,
Phys. Rev. Lett. {\bf105}, 020502 (2010).


\bibitem{Morimae_upload}
T. Morimae, Phys. Rev. A {\bf83}, 042337 (2011).

\bibitem{Morimae_blind}
T. Morimae, V. Dunjko, and E. Kashefi, 
arXiv:1009.3486.

\bibitem{MF}
T. Morimae and K. Fujii, arXiv:1106.3720.

\bibitem{FM}
K. Fujii and T. Morimae, arXiv:1106.3377.

\bibitem{Chen_tricluster}
X. Chen, B. Zeng, Z. Gu, B. Yoshida, and I. L. Chuang, 
Phys. Rev. Lett. {\bf102}, 220501 (2009).

\bibitem{Darmawan}
A. S. Darmawan and S. D. Bartlett,
Phys. Rev. A {\bf82}, 012328 (2010).

\bibitem{experiment1}
R. Kaltenbaek, J. Lavoie, B. Zeng, S. D. Bartlett, and K. J. Resch, 
Nature Phys. {\bf6}, 850 (2010).

\bibitem{experiment2}
W. B. Gao, X. C. Yao, J. M. Cai, H. Lu, 
P. Xu, T. Yang, C. Y. Lu, Y. A. Chen, 
Z. B. Chen, and J. W. Pan, 
Nature Photonics {\bf 5}, 117 (2011).


\bibitem{Landau}
L. D. Landau and E. M. Lifshitz, {\it Statistical Physics},
(Butterworth-Heinemann, Oxford, 1980).

\bibitem{Hall}
D. C. Tsui, H. L. Stormer, and A. C. Gossard, Phys. Rev. Lett.
{\bf48}, 1559 (1982).
\bibitem{Wen}
M. Levin and X. G. Wen, Phys. Rev. B {\bf71}, 045110 (2005).


\bibitem{WenPTP}
X. G. Wen, Prog. Theor. Phys. Suppl. {\bf160}, 351 (2006).
\bibitem{toric1}
E. Dennis, A. Kitaev, A. Landahl, and J. Preskill,
J. Math. Phys. {\bf43}, 4452 (2002).
\bibitem{toric2}
A. Kitaev, Ann. Phys. {\bf303}, 2 (2003).

\bibitem{BR}
S. Bravyi and R. Raussendorf,
Phys. Rev. A {\bf76}, 022304 (2007).

\bibitem{generalize}
The equation above Eq.~(6) in Ref.~\cite{BR} holds 
for the hexagonal lattice.
Any vertex of a subgraph of the two-dimensional square lattice
has degree 1, 2, 3, or 4. They showed that it can be assumed
that all vertices have degree 3.
Since any vertex of a subgraph of the two-dimensional hexagonal lattice
has degree 1, 2, or 3, their result can be applied to the hexagonal lattice.
Thus the discussion around Eq.~(8)
holds for the hexagonal lattice.
Equation (11) holds for the hexagonal lattice.
Lemma on page 5 also holds for the hexagonal lattice.


\bibitem{Miyakeholographic}
A. Miyake, Phys. Rev. Lett. {\bf105}, 040501 (2010).

\bibitem{Bartlett}
S. D. Bartlett, G. K. Brennen, A. Miyake, and J. M. Renes,
Phys. Rev. Lett. {\bf105}, 110502 (2010).

\bibitem{LU}
X. Chen, Z. C. Gu, and X. G. Wen, 
Phys. Rev. B {\bf82}, 155138 (2010).
\bibitem{Yoshida}
B. Yoshida, Ann. Phys. {\bf326}, 15 (2011).
\bibitem{Hastings}
M. B. Hastings, arXiv: 1106.6026.

\bibitem{tensor_representation1}
X. Chen, B. Zeng, Z. C. Gu, I. L. Chuang, and X. G. Wen,
Phys. Rev. B {\bf82}, 165119 (2010).
\bibitem{tensor_representation2}
Z. C. Gu, M. Levin, B. Swingle, and X. G. Wen,
Phys. Rev. B {\bf79}, 085118 (2009).

\bibitem{single_can_be_done}
Universal single-qubit rotation can be also done without 
such an encoding.
However, in this case, the rerouting technique~\cite{Gross1,Gross2,Gross3}
is required.
At this stage, we don't know which is easier: encoding or rerouting.
It should depend on specific experimental setups.

\bibitem{two_encoding}
Two types of logical encoding are required, since,
for example, the operation $I\otimes X$ brings state
outside of one logical encoding space.

\bibitem{finite_size_effect}
When a horizontal wire is decoupled,
its MPS is given by $A[0]=I$ and $A[1]=Z$
or $A[0]=X$ and $A[1]=XZ$.
Thus our resource state is in the class of Ref.~\cite{FM},
and therefore it is free from the finite-size effect~\cite{FM}.










\end{thebibliography}
\end{document}